%% file: main.tex
\documentclass[sigconf, authorversion]{acmart}
\usepackage{listings}
\usepackage{marvosym}
\usepackage{graphicx}
\usepackage{subcaption, multirow, pifont, xspace}
\usepackage[normalem]{ulem}
\useunder{\uline}{\ul}{}
\newcommand{\methodnameshort}{\textsc{PMG}\xspace}
\usepackage[export]{adjustbox}
\usepackage{ifsym}

\AtBeginDocument{%
 }

\copyrightyear{2024}
\acmYear{2024}
\setcopyright{rightsretained}
\acmConference[WWW '24]{Proceedings of the ACM Web Conference 2024}{May 13--17, 2024}{Singapore, Singapore}
\acmBooktitle{Proceedings of the ACM Web Conference 2024 (WWW '24), May 13--17, 2024, Singapore, Singapore}\acmDOI{10.1145/3589334.3645633}
\acmISBN{979-8-4007-0171-9/24/05}

\makeatletter
\gdef\@copyrightpermission{
  \begin{minipage}{0.3\columnwidth}
   \href{https://creativecommons.org/licenses/by/4.0/}{\includegraphics[width=0.90\textwidth]{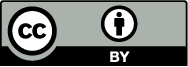}}
  \end{minipage}\hfill
  \begin{minipage}{0.7\columnwidth}
   \href{https://creativecommons.org/licenses/by/4.0/}{This work is licensed under a Creative Commons Attribution International 4.0 License.}
  \end{minipage}
  \vspace{5pt}
}
\makeatother

\begin{document}

\title{PMG : Personalized Multimodal Generation with Large Language Models}

\author{Xiaoteng Shen$^{*\dagger}$}
\orcid{0009-0002-9559-3293}
\affiliation{%
  \institution{Shenzhen International Graduate School, Tsinghua University}
  \city{Shenzhen}
  \country{China}
}
\email{shenxt22@mails.tsinghua.edu.cn}

\author{Rui Zhang$^{*\text{\Letter}}$}
\orcid{0000-0002-8132-6250}
\affiliation{%
  \institution{www.ruizhang.info}
  \city{Shenzhen}
  \country{China}
}
\email{rayteam@yeah.net}

\author{Xiaoyan Zhao$^\dagger$}
\orcid{0000-0001-6001-1260}
\affiliation{%
  \institution{The Chinese University of Hong Kong}
  \city{Hong Kong SAR}
  \country{China}
}
\email{xzhao@se.cuhk.edu.hk}

\author{Jieming Zhu}
\orcid{0000-0002-5666-8320}
\affiliation{%
  \institution{Huawei Noah's Ark Lab}
  \city{Shenzhen}
  \country{China}
}
\email{jiemingzhu@ieee.org}

\author{Xi Xiao$^\text{\Letter}$}
\orcid{0000-0003-1521-9542}
\affiliation{%
  \institution{Shenzhen International Graduate School, Tsinghua University}
  \city{Shenzhen}
  \country{China}
}
\email{xiaox@sz.tsinghua.edu.cn}
\thanks{$^*$Both authors contributed equally to this work.}
\thanks{$^\dagger$The work was done when the authors were interns at Huawei Noah's Ark Lab.}
\thanks{\Letter\ Corresponding authors.}

\renewcommand{\authors}{Xiaoteng Shen, Rui Zhang, Xiaoyan Zhao, Jieming Zhu, and Xi Xiao}
\renewcommand{\shortauthors}{Xiaoteng Shen, Rui Zhang, Xiaoyan Zhao, Jieming Zhu, \& Xi Xiao}

\input{section/abstract}

\input{section/introduction}

\input{section/related_work}

\input{section/method}

\input{section/experiment}

\input{section/conclusion}

\begin{acks}
This work was supported in part by the Overseas Research Cooperation Fund of Tsinghua Shenzhen International Graduate School (HW2021013).
\end{acks}

\bibliographystyle{ACM-Reference-Format}
\balance
\bibliography{sample-base}

\appendix
\input{section/appendix}

\end{document}

%% file: section/abstract.tex
\begin{abstract}

The emergence of large language models (LLMs) has revolutionized the capabilities of text comprehension and generation. Multi-modal generation attracts great attention from both the industry and academia, but there is little work on personalized generation, which has important applications such as recommender systems.
This paper proposes the first method for personalized multimodal generation using LLMs, showcases its applications and validates its performance via an extensive experimental study on two datasets.
The proposed method, Personalized Multimodal Generation (PMG for short) first converts user behaviors (e.g., clicks in recommender systems or conversations with a virtual assistant) into natural language to facilitate LLM understanding and extract user preference descriptions. Such user preferences are then fed into a generator, such as a multimodal LLM or diffusion model, to produce personalized content.
To capture user preferences comprehensively and accurately, we propose to let the LLM output a combination of explicit keywords and implicit embeddings to represent user preferences. Then the combination of keywords and embeddings are used as prompts to condition the generator. We optimize a weighted sum of the accuracy and preference scores so that the generated content has a good balance between them.
Compared to a baseline method without personalization, PMG has a significant improvement on personalization for up to 8\% in terms of LPIPS while retaining the accuracy of generation.


\end{abstract}

\ccsdesc[500]{Information systems~Personalization}
\ccsdesc[500]{Information systems~Multimedia content creation}

\keywords{Multimodal Generation, Large Language Model, Personalization}



\maketitle

%% file: section/introduction.tex
\section{Introduction}
Large language models (LLMs) have demonstrated impressive capabilities in comprehending and generating text. Building upon these achievements, researchers have focused on expanding LLMs into the domain of multimodal understanding, with a particular emphasis on image and audio~\cite{minigpt-4, macaw-llm}. The field of multimodal generation has also gained significant attention, especially following the remarkable video generation capabilities showcased by Sora~\cite{Sora}. To enable multimodal generation tasks, LLMs can be integrated with modality-specific generators such as diffusion models~\cite{diffusion} or multimodal LLMs~\cite{GPT4}.

\begin{figure}
    \centering
    \includegraphics[width=0.8\linewidth]{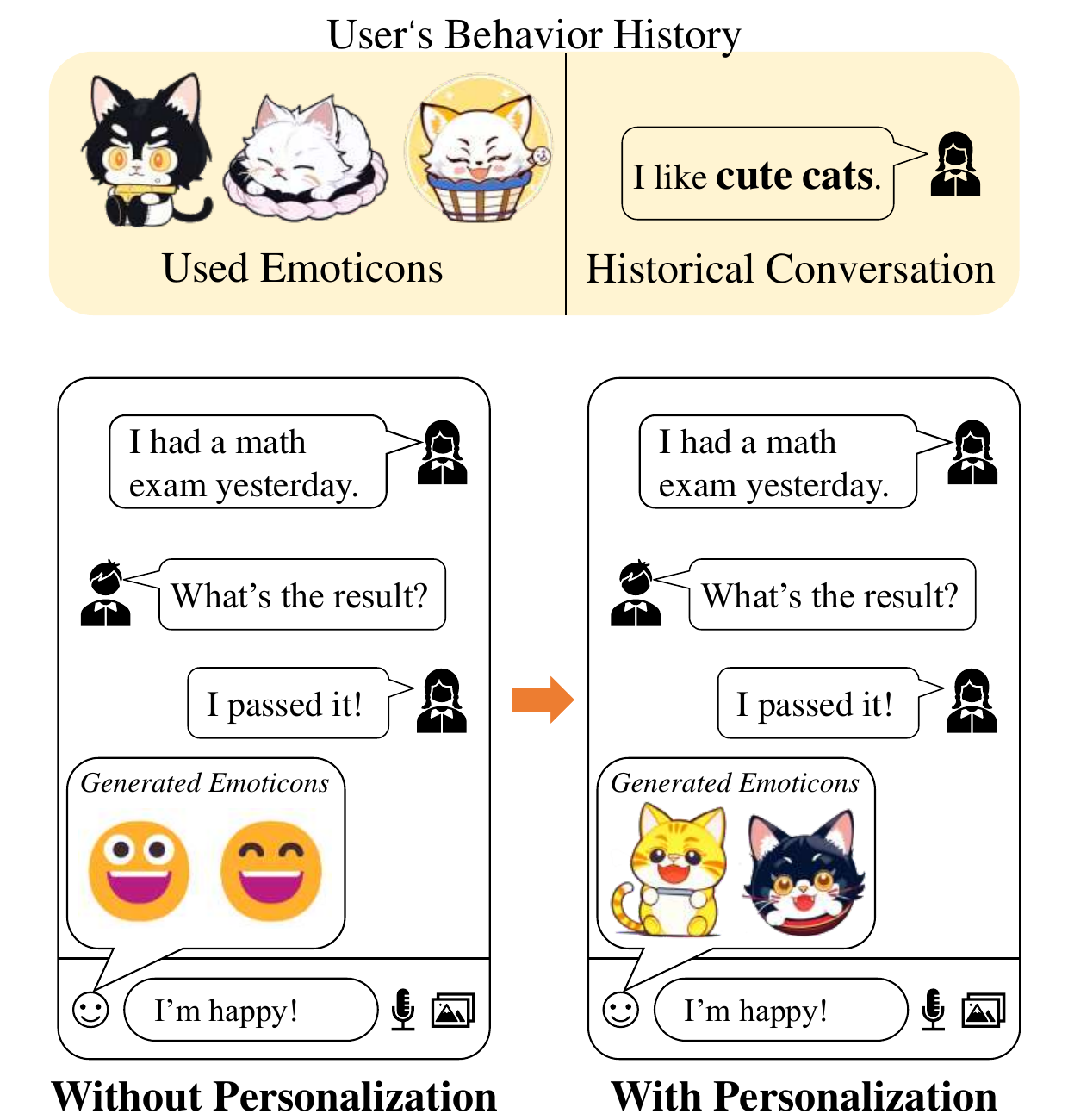}
    \caption{The personalized generation based on user behaviors produces emoticons of a cute cat that are more appealing to cat lovers compared to the normal generation.}
    \label{fig:intro}
\end{figure}

This paper aims to integrate personalization into multimodal generation using LLMs, and to our best knowledge no existing work has addressed this task. Personalization is essential for improving user experience and better meeting users' needs. 
Figure~\ref{fig:intro} shows an example of a chat tool. When the user types in ``I'm happy!'', the chat tool understands the sentiment and automatically recommends emoticons of ``happy'' for the user to choose and click. Popular apps such as TikTok, Discord, WeChat and Telegram already have functions similar to this, but they are without personalization, which is shown in the left part of Figure~\ref{fig:intro}. After adding personalization, the chat tool would be able to generate personalized emoticons that are more appealing to the user as shown in the right part of Figure~\ref{fig:intro}: based on the user's behavior history such as frequently used emoticons (cats in the example) or historical conversation (``I like cute cats'' in the example), the chat tool would generate emoticons of happy cats.

There is a wide range of applications of multimodal generation. For example, online advertisements need well-designed images of products to attract users. When recommending a movie, a personalized generator produces personalized movie posters by amplifying the elements of a movie to the user's preference so that it is more likely to attract the user's attention. Personalized clothing apps can generate images of a person wearing a piece of clothing customized to her preferred height, weight, colors, etc., so the user gets a better idea of what the clothes look like when she wears them. In video games, the background music may be generated to align with the content of the video and the user’s preferred music genre.
Moreover, as the generated content reflects user preferences, they may be leveraged as data augmentation to improve recommendation accuracy.

In the above applications, we refer to the items we aim to generate without personalization as \textit{target items}, e.g., the happy emoticons in the left part of Figure~\ref{fig:intro}; note that there may be multiple target items, e.g., there are multiple smiley faces or multiple candidate movies for recommendation. We refer to the items we aim to generate with personalization as \textit{personalized target items}, e.g., the happy emoticons in the right part of Figure~\ref{fig:intro}. The personalization process should make the candidate target items tuned to users' preferences while retaining their relevance to the candidate target items, where such relevance will be measured by an accuracy score in our experimental study. For example, if we generated a crying cat, the accuracy score would be low in the example of Figure~\ref{fig:intro}.

To address the aforementioned applications, we propose personalized multimodal generation (PMG for short) using LLMs. PMG first extracts a user's preferences from the user's behavior history, such as clicks in recommender systems or past conversations, and converts them into natural language such that they are easily understood by LLMs. The user preferences are then fed into a generator such as a multimodal LLM or diffusion model to condition their generation of the multimodal content. There are a few challenges when implementing our method.

First, we find that merely representing user preferences as natural language, specifically keywords, may not be accurate because they have limited expressive ability whereas user preferences are abstract. To address this challenge, we propose to let the LLM output a combination of explicit keywords and implicit embeddings to represent user preferences. Then the combination of keywords and embeddings are used as prompts to condition the generator.

Second, conditioning the generation process also poses a challenge, as it requires accurately matching both the user preferences and a target item. A naive mixing of these two factors may lead to an imbalance, potentially overshadowing one in the final outcome.
To address this, we employ a weighted sum of the accuracy score and the preference score for each outcome. The accuracy score measures the level of consistency between the generated result and the target item, while the preference score gauges the degree of personalization. We optimize the sum by balancing the weights of the user preferences and target items, allowing us to address the imbalance and customize the degree of personalization.

Our contributions are summarized as follows:
\begin{itemize}
    \item To our knowledge, this is the first work to address the problem of personalized multimodal generation using LLMs, and we demonstrate a wide range of applications.
    \item To address the problem, we propose a method named PMG, which first converts user behaviors into natural language so that LLMs can understand them and extract user preferences. Then the user preferences are fed into a generator to produce personalized content.
    \item To address the challenge of capturing user preferences comprehensively and accurately, we propose to let the LLM output a combination of explicit keywords and implicit embeddings to represent user preferences, which are then used as prompts to condition the multimodal generation. We also propose to optimize a weighted sum of the accuracy score and preference score so that the generated content has a good balance between them.
    \item An extensive experimental study validates the effectiveness of our method. Compared to a baseline method, which does not have personalization, PMG has significant improvement in personalization for up to 8\% in terms of LPIPS while retaining the accuracy of generation.
\end{itemize}

%% file: section/related_work.tex
\section{Related work}

\subsection{Multimodal Generation}
In the field of multimodal generation, previous research has investigated the utilization of generative models like Generative Adversarial Networks (GANs~\cite{GAN}) and Variational Autoencoders (VAEs~\cite{VAE}) to produce diverse and realistic outputs across various modalities. GANs employ a generator network and a discriminator network that undergo adversarial training. On the other hand, VAEs learn latent representations of data and generate new samples. Researchers have extensively explored and enhanced these approaches~\cite{Sketch-RNN, CAN}.

The introduction of CLIP~\cite{clip} revolutionized text-guided generation, making it more accessible. As a result, the diffusion model with CLIP text encoder gained widespread popularity and became the method of choice for various generation tasks, including image generation~\cite{stable_diffusion} and audio generation~\cite{Diffsound}. It is often utilized as a downstream multimodal generator in LLM response generation. While most of these methods~\cite{visual_chatgpt, layoutllm} rely on natural language to establish the connection between the pre-trained LLM and generator, they are hampered by limited natural language expression capability. In contrast, TANGO~\cite{TANGO} and GILL~\cite{gill} employ informative hidden embeddings but are not stable and require substantial training to align their embedding space.

The current approaches to personalized generation, such as Textual Inversion~\cite{textual_inversion} and DreamBooth~\cite{dreambooth}, mainly focus on integrating new characters or image styles into a pre-trained diffusion model using a few images. These approaches differ significantly from personalization based on user behaviors, which emphasizes the general interests of users rather than specific instances. Moreover, user behaviors encompass a combination of clicked items (including textual and visual features), conversations, and more, making it impractical to process using existing personalized generation.

\subsection{LLM for Recommendation}
Recommendation~\cite{stem} is an important means for information retrieval and many studies aim to leverage the exceptional reasoning capabilities of LLMs for recommender systems. The predominant approaches utilize the textual feature of items in historical click sequences and candidate pools so that the LLM can directly generate recommended items. Although it can yield favorable results even without training~\cite{LLMRanker, Chatrec, NIR}, this approach lacks specific optimization for recommender tasks. Certain studies~\cite{UniCRS, m6rec, tallrec} follow this paradigm but employ techniques like prompt learning~\cite{L2P} or LoRA~\cite{lora} for fine-tuning the LLM and enhancing recommendation accuracy. On the other hand, P5~\cite{P5} primarily utilizes ID features rather than textual features to cater to recommendation tasks.

As for multimodal recommendation, VIP5~\cite{vip5} builds upon P5 by incorporating item images as visual features and introduces adapters to understand them. MISSRec~\cite{MISSRec} is a pre-training method for multimodal sequential recommendation, which focuses on learning universal item representations with multimodal features. However, the above methods only have multimodal understanding ability but not multimodal generation ability, i.e., the recommended items by these methods will have images only if those images are already available in the items database; if an item does not have any image available, these methods cannot generate one when recommending the item.

%% file: section/method.tex
\section{Method}
\subsection{Overview}
\begin{figure*}
    \centering
    \begin{minipage}[b]{0.8\linewidth}
        \centering
        \centerline{\includegraphics[width=\linewidth]{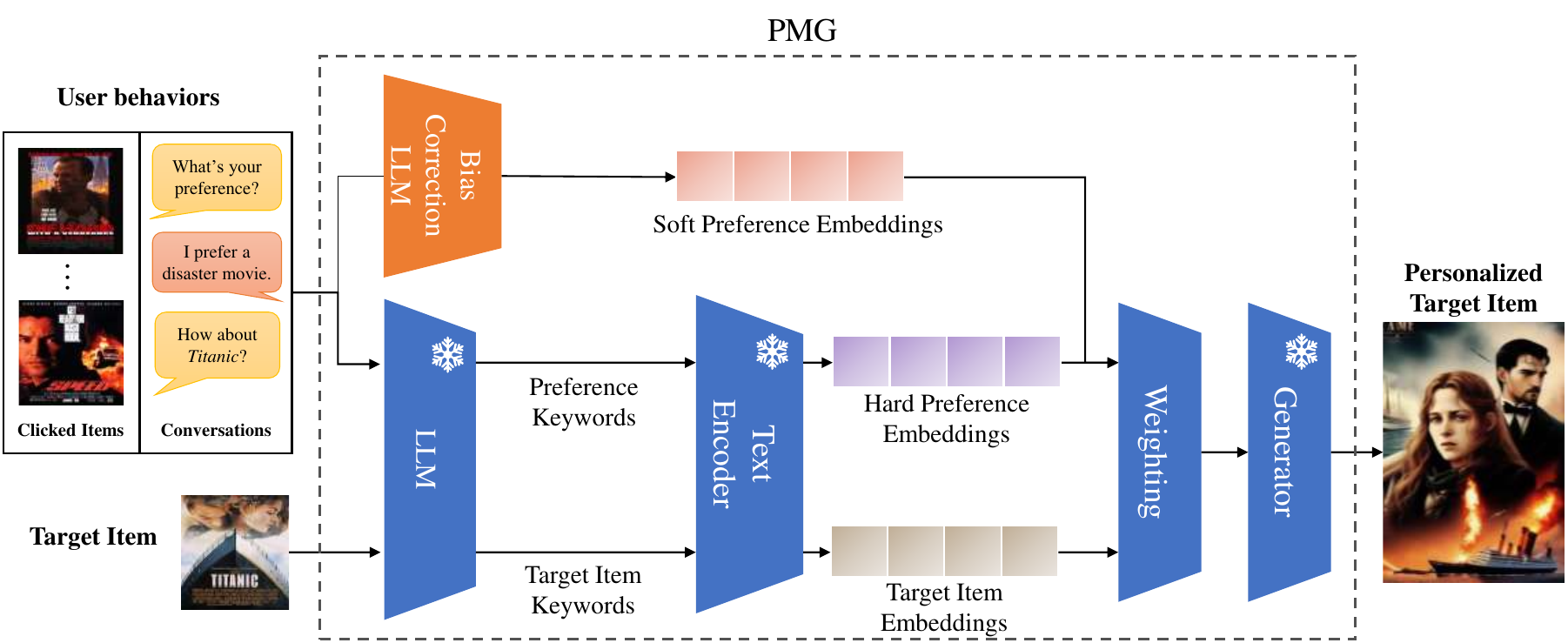}}
    \end{minipage}
    \caption{Overview of our method. By utilizing user behaviors and a target item as input, we generate personalized multimodal content for the item, taking a movie poster as an example in the figure.}
    \label{fig:overview}
\end{figure*}

Our proposed method \methodnameshort is depicted in Figure~\ref{fig:overview}. We leverage the reasoning abilities of an LLM to extract user preferences from historical behaviors (including clicks in recommender systems and conversations with a virtual assistant). The user behaviors are used to produce preference conditions, including explicit keywords in natural language (named preference keywords) by a frozen LLM and implicit embeddings (named soft preference embeddings) by a tuned LLM for multimodal bias correction~\cite{gill}. Additionally, we convert the target item into explicit keywords (named target item keywords) to serve as the target item conditions. Ultimately, the generator, which could be a diffusion model or multimodal LLM, produces the results by incorporating and weighting preference and target item conditions after the text encoder of the generator.

\subsection{Generate Explicit Keywords}
\label{sec:Construct_prompt}
Given our objective of extracting user preferences using an LLM from behaviors, the simplest and most effective approach is to convert user behaviors into text and analyze them using the LLM. The generator typically has a limited input length (e.g., 77 tokens in Stable Diffusion~\cite{stable_diffusion}), making keyword summarization more informative than using full sentences. As a result, we design prompts for each scenario and leverage the zero-shot capability of the LLM without the need for training. In the following, we will discuss the process of prompt design.

\subsubsection{\textbf{Preprocess of user behaviors.}}
We consider two types of user behaviors: historical clicks $ H = \left\{h_1, h_2, \cdots\right\}$ and conversations $ C = \left\{c_1, c_2, \cdots\right\}$.
The input features could be multimodal, including texts, images, audios, etc. Normally, the LLM has the ability to handle complex texts, so we can simply feed the texts into it. But the texts may be long (e.g., a plot synopsis of a movie), and concatenating all of them from an item sequence exceeds the token length limit of the LLM. In this case, we summarize the text features of each item and conversation into a short sentence using the LLM as preprocessing. For the other features, we convert them into text using a caption model (e.g. BLIP-2~\cite{blip-2}, CLAP~\cite{clap}) or using multimodal LLM (e.g. MiniGPT-4~\cite{minigpt-4}, mPLUG-owl~\cite{mplugowl}) capable of processing multimodal inputs. The purpose of this preprocessing is to summarize the features, reducing redundancy and preserving long-term contexts. Formally, this process can be defined as follows:
\begin{equation}
\label{eq:summarization}
\nonumber
\begin{aligned}
  x_{i} &= \left[LLM_g(t_{h_i}), LLM_g(v_{h_i}), \cdots\right], \\
  y_{i} &= \left[LLM_g(t_{c_i}), LLM_g(v_{c_i}), \cdots\right],
\end{aligned}
\end{equation}
where $ t, v, \cdots $ are textual, visual and other multimodal features, $ x_i $ and $ y_i $ denote the summarized data of historical items and conversations. $ LLM_g $ represents the generating operation of LLM, distinguishing from its forward operation $LLM_f$.

\subsubsection{\textbf{Construction of prompt.}}
Using the behavior information $ \mathbf{x}, \mathbf{y} $, we can construct a prompt to extract user preferences with the help of the LLM. There are three additional components: the instruction principle $ p $, attribute $ a_i $, and examples $ e $. These components are artificially designed for each scene. The principle $p$ describes the task being performed by the LLM, which is ``user preference extraction''. The attributes $\mathbf{a}$ are tailored for each scene, such as ``color, material, shape'' for clothes or ``genre, director, origin'' for movies. In each question, LLM is assigned the task of answering user preferences related to a specific attribute, and the answers are later combined. The examples $e$, which provide the desired output format and example keywords (e.g., ``cute'', ``cartoon'', etc.), not only assist in guiding the LLM's responses but also follow a standardized output format, thereby facilitating the extraction of keywords from the generated output. Using this prompt, we can represent the keywords $ \mathbf{k}^p_i $ generated by LLM for attribute $ a_i $ as follows:
\begin{equation}
\nonumber
\begin{aligned}
  \mathbf{k}^p_i &= LLM_g\left(p, a_i, e, \mathbf{x}, \mathbf{y}\right).
\end{aligned}
\end{equation}
Next, we combine the outputs of each attribute and eliminate any duplicates to obtain preference keywords $ \mathbf{k}^p $. The process of generating the target item keywords $\mathbf{k}^t$ is similar but with only one target item $ h^t $ and its corresponding summarized information $ x^t $. In this case, there are no conversations involved, and there is only an overall attribute (the union of all the above attributes):
\begin{equation}
\nonumber
\begin{aligned}
  \mathbf{k}^t &= LLM_g\left(p, e, x^t\right).
\end{aligned}
\end{equation}

\subsection{Generate Soft Preference Embeddings}
\begin{figure}[t]
    \centering
    \includegraphics[width=0.9\linewidth]{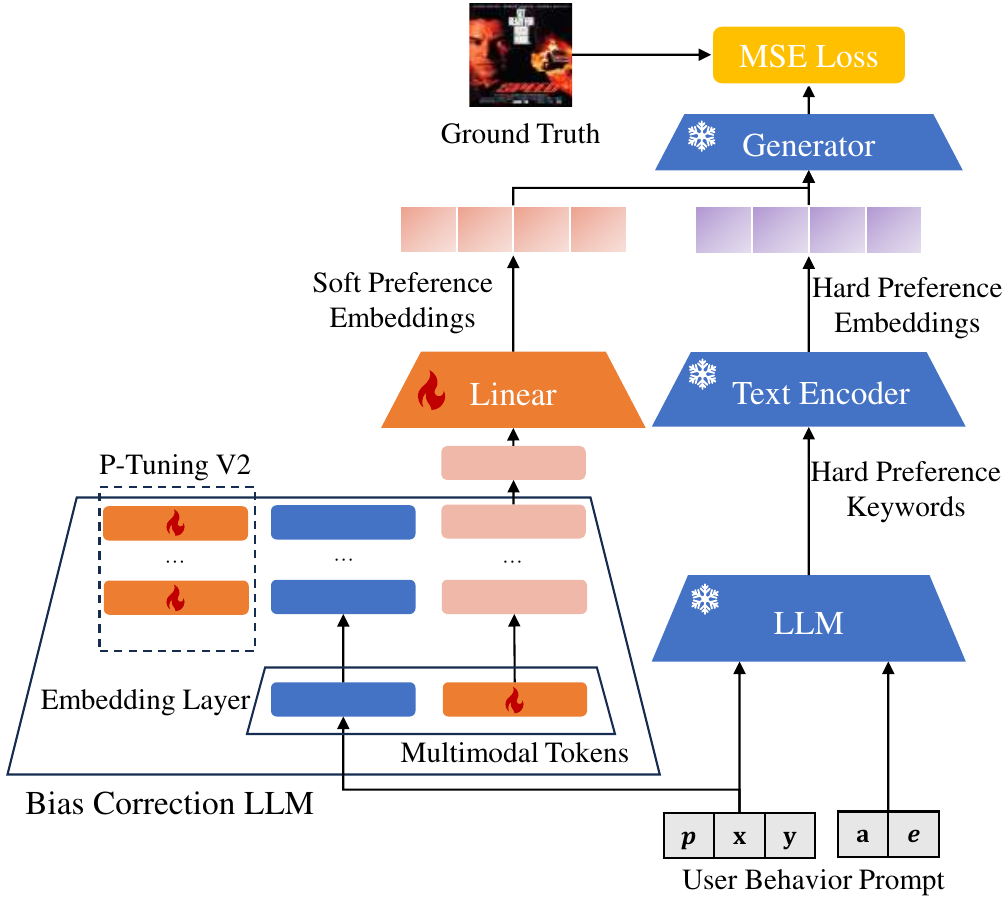}
    \caption{Model designed to train soft preference embeddings.}
    \label{fig:model}
\end{figure}

We have developed a method that relies solely on explicit keywords for representation. However, natural language, as a discretized form, has limited expressive capabilities with limited length. On the other hand, utilizing continuous hidden embeddings, which offer more informative and precise representations, requires substantial training resources. We utilize natural language as the baseline while training soft preference embeddings as an extra signal to correct this language bias with the help of an LLM, named bias correction LLM. These embeddings assist in addressing the mismatch between the natural language baseline and the actual user interests. The model is illustrated in Figure~\ref{fig:model}.

\subsubsection{\textbf{Bias Correction LLM}}
The primary objective of the LLM is to predict the next textual token so it can only understand and generate texts. However, when applied to multimodal generation, it becomes necessary to introduce multimodal tokens to acquire the ability of multimodal generation. Inspired by GILL\cite{gill}, we incorporate multimodal tokens as learnable parameters into the embedding table and then utilize a linear layer to align the embedding space of the LLM with that of the generator. This alignment ensures consistency and compatibility between the LLM and the text encoder of the generator, facilitating the generation process. Additionally, we employ P-Tuning V2~\cite{P-tuningv2} to fine-tune the LLM specifically for the generation task, which can enhance its generation ability. During each inference, the multimodal tokens are appended after the user behavior prompt. The soft preference embeddings are obtained by passing these augmented inputs through the LLM (with P-Tuning V2) and the linear layer.

Formally, in conjunction with the user behavior prompt $ p, \mathbf{x}, \mathbf{y} $ constructed in section~\ref{sec:Construct_prompt}, we include additional multimodal tokens $ \mathbf{m} = \left\{m_1, \cdots, m_L\right\} $ of length $ L $. Attributes and examples are not utilized in this context, as the prefix embeddings have the ability to learn them on their own. These tokens are passed to the LLM, and their corresponding embeddings in the embedding layer are trainable. Following the P-Tuning V2 approach, $ S $ trainable prefix embeddings $ \mathbf{t} = \left\{t_1, \cdots, t_S\right\} $ are prepended to the embedding sequence in the self-attention of each transformer layer. The resulting output embeddings in the LLM's forward operation can be represented as:
\begin{equation}
\label{eq:LLM_embedding}
\nonumber
\begin{aligned}
  \mathbf{prompt} &= \left(p, \mathbf{x}, \mathbf{y}\right), \\
  \left[ \mathbf{E}_{prompt}, \mathbf{E}_m\right] &= LLM_f\left(\mathbf{t}, \mathbf{prompt}, \mathbf{m}\right),
\end{aligned}
\end{equation}
where $ \mathbf{E}_{prompt}, \mathbf{E}_m $ represent the output embedding of LLM, and the soft preference embeddings $ \mathbf{E}_m $ is used for the subsequent multimodal generation process.

\subsubsection{\textbf{Training with multimodal supervision.}}
In contrast to GILL~\cite{gill}, which solely relies on captions for supervision, we believe that incorporating multimodal supervision (such as real images or audios) is more meaningful and helps to correct deviations. However, this approach introduces the challenge of propagating gradients backward through the generator, resulting in increased training difficulty. To simplify training, we utilize the preference keywords generated in section~\ref{sec:Construct_prompt} as a foundational framework and focus on training a limited number of soft preference embeddings as additional conditions for the generation process.

The preference keywords are tokenized and transformed into hard preference embedding $\mathbf{E}_k$ by the text encoder of the generator. Then, we concatenate the $\mathbf{E}_m$ and $\mathbf{E}_k$ as conditioning input for the generator. Regarding data splitting, since it is impossible to obtain a real personalized image as ground truth, we use the last item in the interaction sequence as supervision and the others as input.

Different generator models have different training algorithms. In our implementation, we utilize a diffusion model, which contains a text encoder and a U-Net~\cite{unet}. The U-Net is employed as a conditional denoising module to generate images through multiple denoising steps. Following its training process, we introduce random noise $\epsilon \sim \mathcal{N}(0, 1)$ to the multimodal supervision $M_s$ and then attempt to denoise it:
\begin{equation}
\nonumber
\begin{aligned}
    \mathbf{E}^p &= concatenate(\mathbf{E}_m, \mathbf{E}_k) \\
    M_n &= M_s + \epsilon, \\
    M_d &= Unet(\mathbf{E}^p, M_n).
\end{aligned}
\end{equation}
The loss is calculated as MSE loss of $ M_s $ and $ M_d $:
\begin{equation}
\nonumber
\begin{aligned}
    loss &= MSE(M_s, M_d).
\end{aligned}
\end{equation}
Using this loss, we train the embeddings of multimodal tokens, and prefix embeddings in P-Tuning v2 to enable the multimodal generation ability of LLM, together with the mapper layer to align embedding space.


\subsection{Balancing the accuracy score and the preference score}
\label{sec:re-weighting}

Different from the training process of soft preference embeddings including only preference conditions, the generation inference process incorporates both preference and target item conditions. Simply combining these conditions can result in favoritism towards one and overshadowing the other.
Following previous studies such as DreamBooth~\cite{dreambooth} and GILL~\cite{gill}, we use the similarity between the generated results and the preference keywords to measure the degree of personalization, which we call the \textit{preference score}, and the \textit{accuracy score} refers to the similarity with the target item keywords.
The accuracy score measures the level of consistency with the target item, while the preference score about preference conditions gauges the degree of personalization. To balance them, we employ a weighted sum of accuracy score and preference score using pre-trained multimodal networks (e.g., CLIP~\cite{clip}, CLAP~\cite{clap}).

Assuming the multimodal result $ M $ is generated by:
\begin{equation}
\nonumber
\begin{aligned}
    M &= Generator(w_p\cdot\mathbf{E}^p, w_t\cdot\mathbf{E}^t),
\end{aligned}
\end{equation}
where $ w_p $, $ w_t $ are weights of preference and target item conditions to be adjusted.
Through the encoders of the pre-trained multimodal network, we can transform the result $ M $ and keywords $ \mathbf{k}^p, \mathbf{k}^t $ into embeddings $ e_M, e_p, e_t $. Then we can calculate the similarity between them as the preference score $ d_p $ and accuracy score $ d_t $.
\begin{equation}
\nonumber
\begin{aligned}
    d_p &= \frac{e_M\cdot e_p}{\left\| e_M \right\|_2\left\| e_p \right\|_2}, \\
    d_t &= \frac{e_M\cdot e_t}{\left\| e_M \right\|_2\left\| e_t \right\|_2}.
\end{aligned}
\end{equation}
Finally, our objective is to optimize the weighted sum of $ d_p $ and $ d_t $.
\begin{equation}
\nonumber
\begin{aligned}
    z &= \alpha\cdot \log{d_p} + (1-\alpha)\cdot \log{d_t}.
\end{aligned}
\label{eq:re-weighting}
\end{equation}
The hyper-parameter $ \alpha $ is normally $ 0.5 $ and can be adjusted to achieve different effects according to usage scenarios and needs.

Considering the powerful parallel generation ability of current multimodal generators, we generate with multiple predefined sets of weights $ w_p, w_t $ and pick the one with the highest score $ z $.

%% file: section/experiment.tex
\section{Experiment}
Our method can be used to generate various multimodal content, encompassing not only images and audios but also other modalities. In this section, we focus on the generation of images as it is considered the most common and intuitive modality.
Please consult Appendix \ref{sec:Implement} for the code and implementation details.
Our experiments aim to answer the following research questions:
\begin{itemize}
    \item \textbf{RQ1}: Can \methodnameshort accurately generate images that combine user preferences?
    \item \textbf{RQ2}: Why is conditions weighting necessary?
    \item \textbf{RQ3}: How do explicit keywords and implicit embeddings impact performance?
    \item \textbf{RQ4}: Are P-Tuning v2 and multimodal tokens beneficial while training soft preference embeddings?
    \item \textbf{RQ5}: Are there any additional purposes or applications for the generated images beyond user display?
\end{itemize}

\subsection{Experimental Setup}
\subsubsection{\textbf{Scenarios and dataset.}}
We design the following three scenarios to verify our method:

(1) \textbf{Generating personalized images} of products whose original images are missing according to the historically clicked products of the user. We adopt POG~\cite{pog}, a multimodal dataset of fashion clothes, for training and evaluation. We selected 2,000 users and 16,100 items for experiments.

(2) \textbf{Generating personalized posters of movies} according to historical watched movies of user. We adopt the small version of MovieLens Latest Datasets~\cite{movielens}, which contains 9,000 movies, 600 users, and 100,000 rating interactions.

(3) \textbf{Generating emoticons} in instant messaging according to current conversation and historically used emoticons of the user. Since we cannot find a suitable dataset, we do not train soft preference embeddings and only use keywords to generate images.

The datasets themselves don't include conversations, so we designed some templates to construct them.

\subsubsection{\textbf{Evaluation metrics.}}
We employ multiple image similarity metrics to assess the resemblance between the generated image and historical/target items, quantifying the level of visual personalization achieved. To prevent potential information leakage, we exclude the CLIP metric used in the weighting module from this evaluation. Instead, we utilize the following two metrics:

\textbf{(1) LPIPS} (Learned Perceptual Image Patch Similarity)~\cite{LPIPS}: This metric measures the perceptual similarity between two images by considering human visual perception. It focuses on capturing semantic information.

\textbf{(2) SSIM} (Structural Similarity Index Measure)~\cite{SSIM}: Widely used in image similarity assessment, this metric considers luminance, contrast, and structural information. It places more emphasis on image quality.

By employing these metrics, we can comprehensively evaluate the visual similarity between the generated image and the historical/target items, providing insights into the effectiveness of our personalized generation approach.
Furthermore, we also conduct a human evaluation to verify its effectiveness in the real-world. 

\subsection{Image Comparison (RQ1)}
\label{sec:exp-baseline}

\begin{figure}[t]
    \centering
    \includegraphics[width=0.96\linewidth]{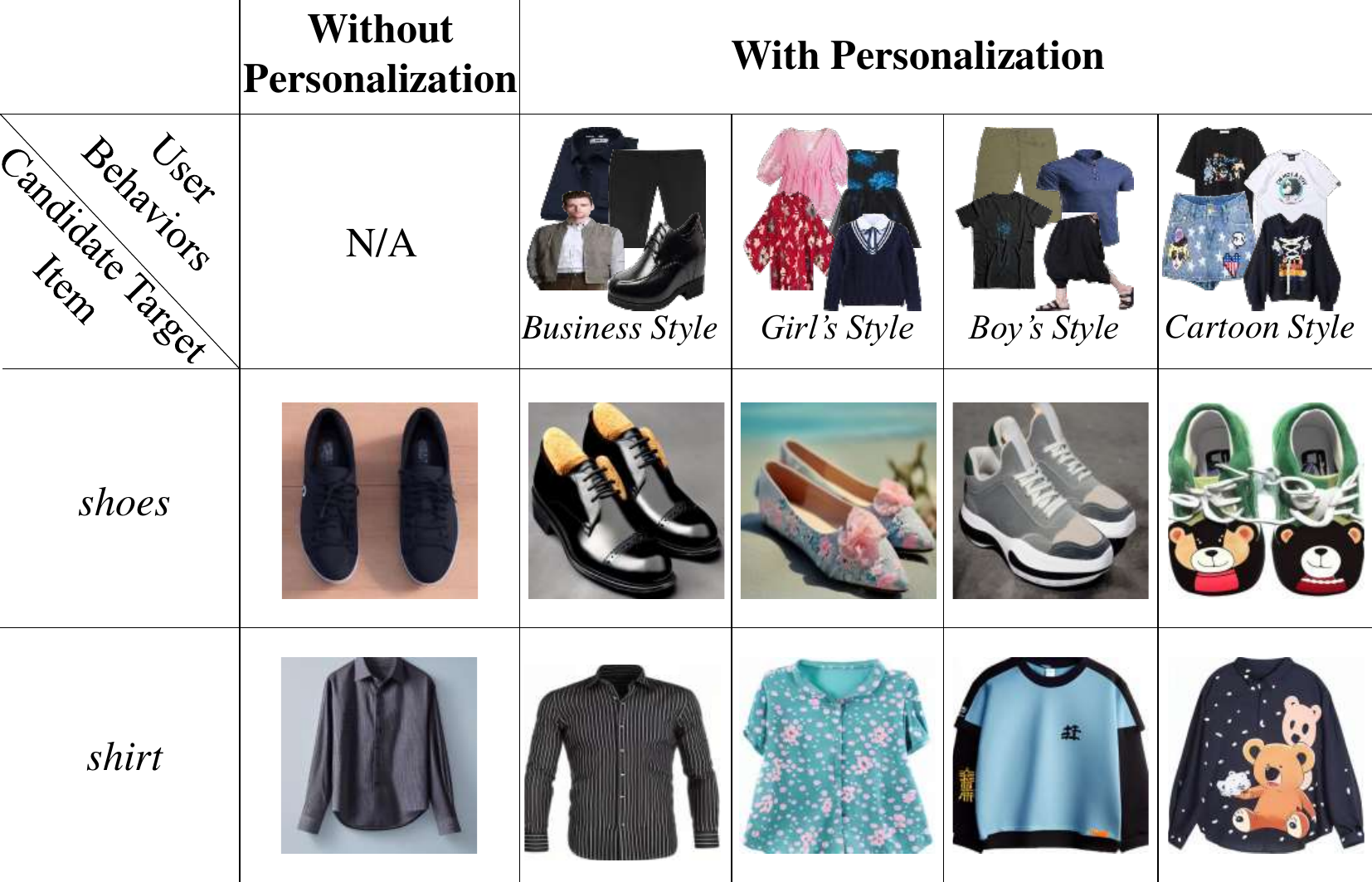}
    \caption{Generated image comparison of our method \methodnameshort in the costume scene. Four typical users with different styles of historical items are picked as input to generate images of shoes and a shirt.}
    \label{fig:baseline-costume}
\end{figure}

\begin{figure}[t]
    \centering
    \includegraphics[width=0.91\linewidth]{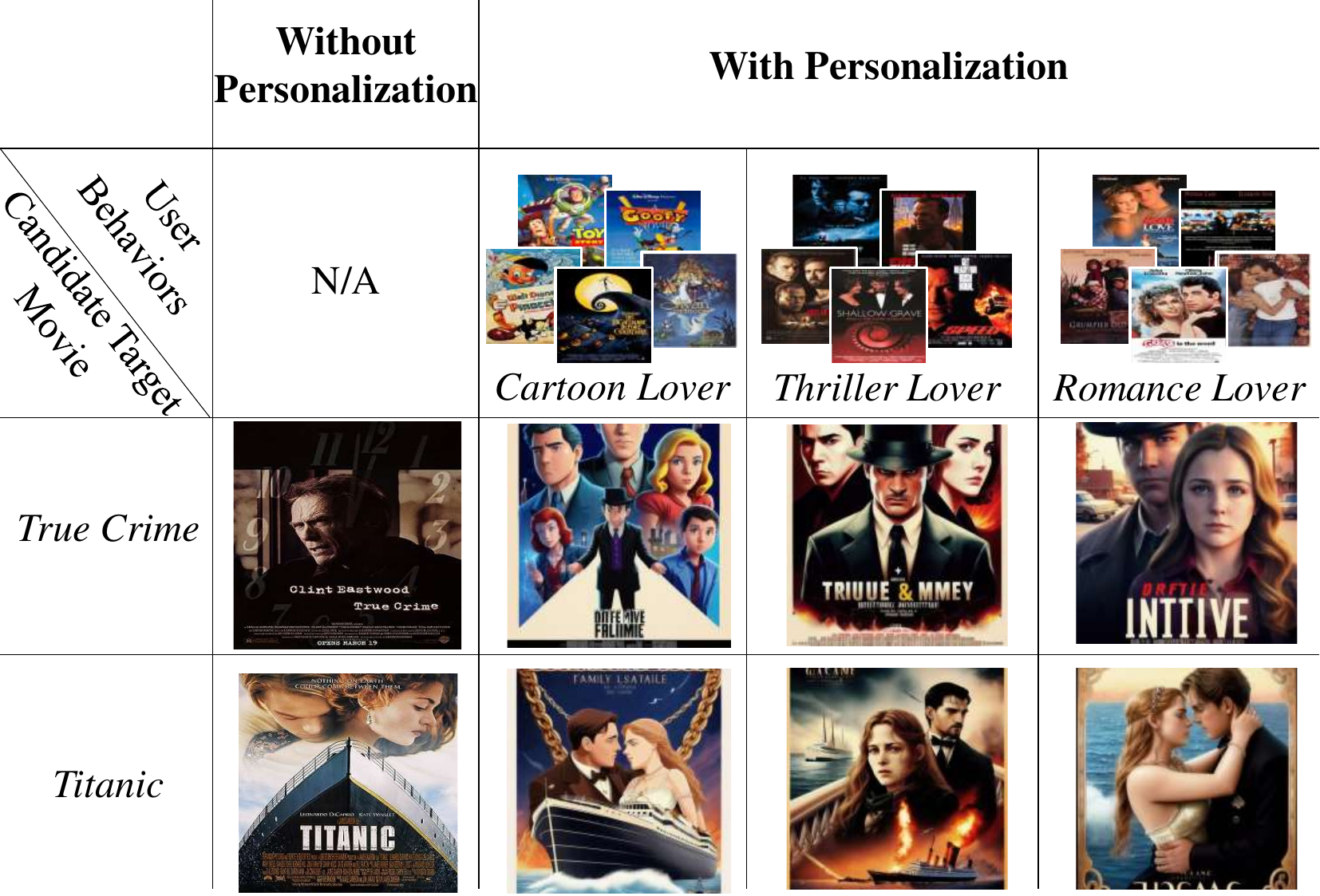}
    \caption{Generated image comparison of our method \methodnameshort in the movie poster scene. Three users with different movie interests are picked as input to generate posters of movie \textit{True Crime} and \textit{Titanic}.}
    \label{fig:baseline-poster}
\end{figure}

\begin{figure}[t]
    \centering
    \includegraphics[width=0.82\linewidth]{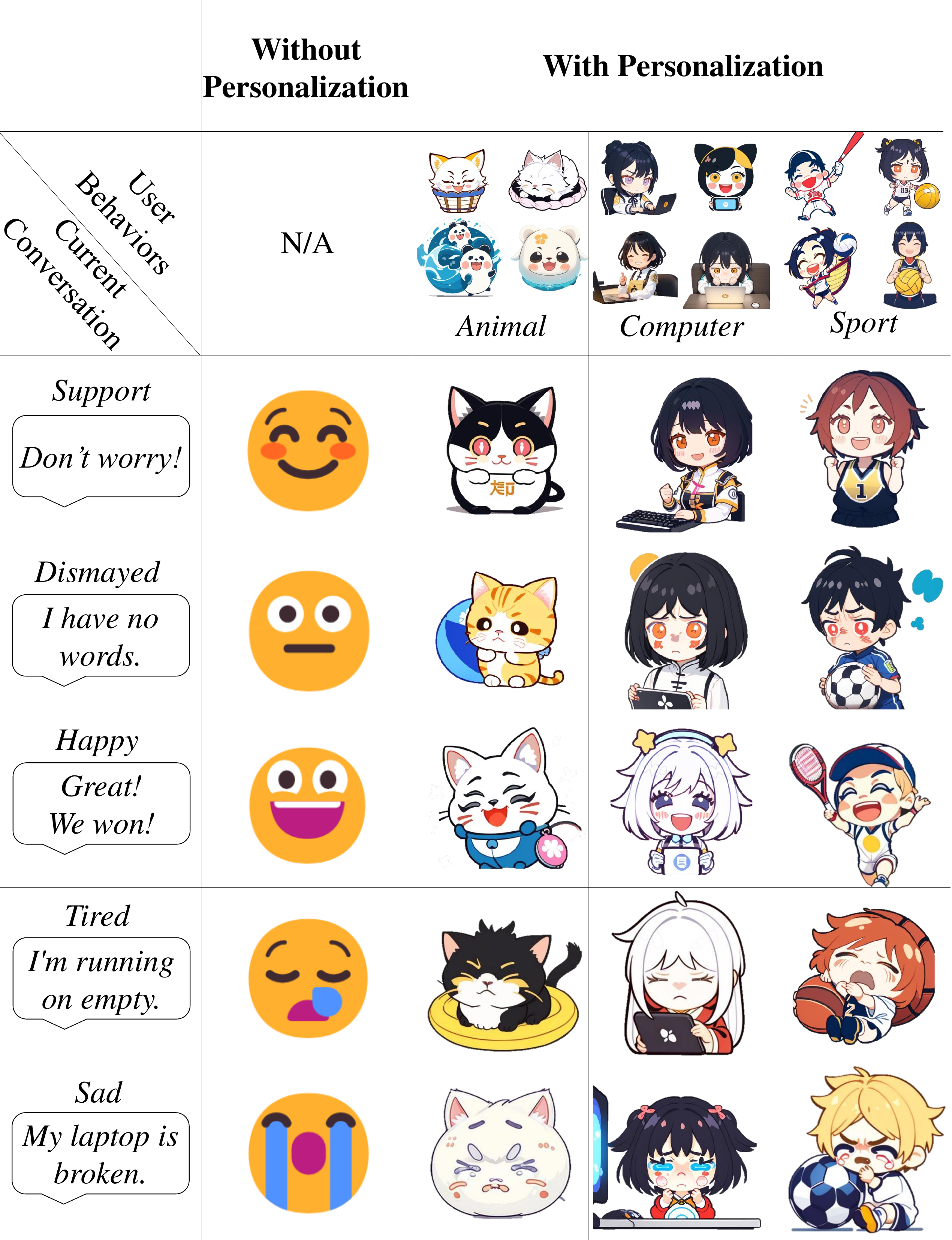}
    \caption{Generated image of our method \methodnameshort in the emoticon scene. We generated emoticons for three users who have different interests in five different emotional conversations. The current conversation serves as the target item.}
    \label{fig:baseline-emoticon}
\end{figure}

\input{img/weight}

In this section, we show the generated images in three scenes: the costume scene, the movie poster scene and the emoticon scene. The existing personalization generation methods such as Textual Inversion~\cite{textual_inversion} and DreamBooth~\cite{dreambooth} train extra embeddings for each user using their historical item images. They are only suitable for scenarios with a small number of users as they can consume significant training resources. As a result, they are not used in our experiments as baselines.

In the costume scene (Figure~\ref{fig:baseline-costume}), \methodnameshort demonstrates notable personalization capabilities, particularly in cartoon and girl's styles. In the cartoon style, \methodnameshort identifies the association of these items with a specific cartoon character and accordingly selects a cartoon bear as the generated output. In the girl's style, \methodnameshort incorporates numerous floral patterns that align with girls' preferences.

In the movie poster scene (Figure~\ref{fig:baseline-poster}), \methodnameshort adeptly combines user preferences with the target item. For instance, in the thriller movie \textit{True Crime}, \methodnameshort consistently incorporates crime and horror elements into the generated posters, regardless of the user generating them. In the case of the romance movie \textit{Titanic}, the generated posters consistently feature a couple in love, while the styles vary based on user preferences. 

In the emoticon scene (Figure~\ref{fig:baseline-emoticon}), we generate emoticons based on the ongoing conversation and previously used emoticons. Utilizing historical emoticons, the LLM helps summarize the user's preferences and designs cartoon characters like cats or football-playing boys.
Then, the LLM analyzes the conversation to identify its emotion and devises a suitable pose for the emoticon, such as crying sadly or squinting from fatigue. Finally, the character and the pose can be considered as the preference condition and target condition respectively to generate the final emoticon.
As a result, we generate emoticons featuring a cat for animal lovers and emoticons relating to balls for sports enthusiasts, among others, and the emotions conveyed are generally accurate.

However, PMG is unable to generate images consistent with real entities. For example, the characters in the generated movie posters may not match the real actors, and the clothing may not match the real products. We will discuss and improve it in future work.

\subsection{Human Evaluation (RQ1)}
The image comparison based on image similarity metrics demonstrates the personalization of generated images, but it cannot be determined whether they can attract users in real-world scenarios. To address it, we conduct a human evaluation to compare the images generated by our method PMG, Textual Inversion~\cite{textual_inversion}, and images without personalization. In Textual Inversion, we use only the images of historically clicked items to learn user preference. We invited 40 volunteers to score 60 images (20 images of each kind) from 1 to 3 in two scenarios (higher scores mean better results). The average scores given by the volunteers are in Table~\ref{tab:human_evaluation}.

\input{table/human_evaluation}

As we observe from the human evaluation result, our method PMG, which is based on multi-modal user behavior outperforms Textual Inversion which is based on only historical clicked images. The human evaluation validates the effectiveness of PMG.

\subsection{Case Study (RQ2)}
As explained in Section \ref{sec:re-weighting}, directly combining personalization and target conditions can result in an imbalance. In Figure \ref{fig:re-weighting}, we observe variations in the generated poster while adjusting condition weights for a romantic target movie \textit{Titanic} and a disaster enthusiast. When the condition weights are set to $w_p:w_t=0:4$, the poster predominantly considers the target condition (romance) and depicts a couple in love. Conversely, when the weights are adjusted to $w_p:w_t=4:0$, the poster focuses solely on the preference condition (disaster) and portrays a ship in a storm.

In order to incorporate both romance and disaster while following our selection principle outlined in Equation \ref{eq:re-weighting}, we evaluate the generated posters based on their $z$ scores. Figure \ref{fig:re-weighting-chosen} achieves the highest $z$ score and is selected as the final output.

\subsection{Ablation Study}

\subsubsection{\textbf{preference conditions. (RQ3)}}
\input{table/ablation_hardsoft}
In this section, we examine the contribution of the two forms of user preference representation, preference keywords, and soft preference embeddings (Table~\ref{tab:ablation_hardsoft}). By calculating the similarity between generated images and historical items, we can measure the degree of personalization, and by calculating the similarity with the target item, we can ensure that our generation does not deviate from the target.

Our method incorporates user preferences, reflected in historical items, and surprisingly, the similarity with the target item even increased in movie scenes. This demonstrates that personalization can smooth out errors between the generator and real scenes. Keywords greatly enhance similarity in both LPIPS and SSIM metrics, while soft preference embeddings reduce LPIPS but not SSIM. This indicates that embeddings introduce personalized semantic information but don't improve image quality due to instability. By combining preference keywords and soft preference embeddings, we achieve rich personalized content without deviating from the target items, while ensuring image quality.

\input{img/soft_prompt}
Figure~\ref{fig:embedding} is a case study on the soft preference embeddings. When provided with only the keywords "shoes, cartoon", there is a certain probability of generating cartoon-style drawings of shoes. However, after incorporating the soft preference embedding, the model consistently generates realistic shoes adorned with cartoon patterns.

\subsubsection{\textbf{Prompt tuning. (RQ4)}}
\input{table/ablation_finetune}
In this section, we analyze the impact of P-tuning V2 and multimodal tokens on the degree of personalization, measured by LPIPS similarity between generated images and historical items. Table~\ref{tab:ablation_finetune} showcases their effectiveness. P-tuning V2 greatly enhances the ability of LLM to extract user preferences. Similarly, multimodal tokens exhibit a positive effect, although they also occupy a limited condition embedding and reduce the number of effective keywords. Therefore, the number of multimodal tokens should not be large, and setting $L=4$ or $L=8$ is determined to be the optimal parameter.

\subsection{Auxiliary Generation (RQ5)}
\input{table/auxiliary}
Our approach extensively explores interest modeling with LLM, enabling the generation of images that can be utilized not only for displaying to users but also for downstream recommendation tasks. This section presents an experiment on the MovieLens, aiming to evaluate the impact of incorporating generated images as additional visual features. To perform the evaluation, we employ MMGCN~\cite{mmgcn} as the base multi-modal recommendation model.

The MovieLens dataset inherently includes image features of items, specifically the original movie posters, but it lacks image features for users. As a result, we have designed the following experiments:
(1) \textbf{No-image}: This experiment does not utilize any image features and relies solely on the IDs of items and users.
(2) \textbf{Item-only}: This experiment solely utilizes the image features of items.
(3) \textbf{Averaged-user}: In addition to item image features, user image features are initialized as the average of historically watched items.
(4) \textbf{Generated-user}: In addition to item image features, user image features are initialized as the image generated by \methodnameshort.
It is important to note that the generated images are created under the preference conditions, without a target item.

Table~\ref{tab:auxiliary} provides compelling evidence that the inclusion of image features for items or users significantly enhances recommendation accuracy. Notably, incorporating the images generated by \methodnameshort yields superior results compared to the simple average baseline. These findings underscore the effectiveness of our approach in capturing user interests by leveraging the reasoning capability of LLM. By incorporating the generated images, our method successfully captures and incorporates nuanced user preferences, leading to improved recommendation performance.


%% file: img/weight.tex
\begin{figure*}[t]
  \centering
  \begin{minipage}[b]{0.15\textwidth}
    \centering
    \includegraphics[width=0.8\textwidth]{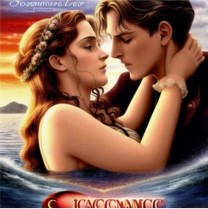}
    \subcaption{$w_p:w_t = 0:4$}
  \end{minipage}
  \hfill
  \begin{minipage}[b]{0.15\textwidth}
    \centering
    \includegraphics[width=0.8\textwidth]{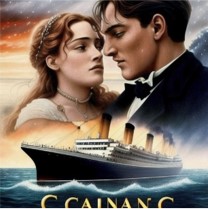}
    \subcaption{$w_p:w_t = 1:3$}
    \label{fig:re-weighting-chosen}
  \end{minipage}
  \hfill
  \begin{minipage}[b]{0.15\textwidth}
    \centering
    \includegraphics[width=0.8\textwidth]{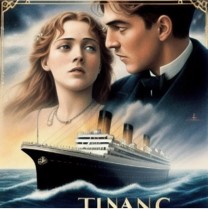}
    \subcaption{$w_p:w_t = 2:2$}
  \end{minipage}
  \hfill
  \begin{minipage}[b]{0.15\textwidth}
    \centering
    \includegraphics[width=0.8\textwidth]{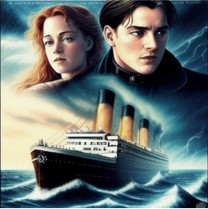}
    \subcaption{$w_p:w_t = 3:1$}
  \end{minipage}
  \hfill
  \begin{minipage}[b]{0.15\textwidth}
    \centering
    \includegraphics[width=0.8\textwidth]{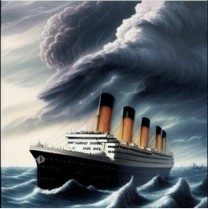}
    \subcaption{$w_p:w_t = 4:0$}
  \end{minipage}
  \caption{Generated poster of movie \textit{Titanic} with different weights of conditions. $w_p$ is the weight of preference conditions, which prefer disaster movie. $w_t$ is the weight of target item conditions, which consider it as a romantic movie. When $w_p:w_t = 1:3$ it achieves the highest $z$ score and the generated poster is a combination of romance and disaster.}
  \label{fig:re-weighting}
\end{figure*}

%% file: table/human_evaluation.tex
\begin{table}
\caption{The average score of generated images in human evaluation}
\begin{tabular}{lcc}
\toprule
                   & Movie Posters Scenario & Clothes Scenario \\ \hline
PMG                & 2.587                  & 2.001            \\
Textual Inversion  & 1.952                  & 1.725            \\
No personalization & 1.462                  & 1.495            \\
\bottomrule
\end{tabular}
\label{tab:human_evaluation}
\end{table}

%% file: table/ablation_hardsoft.tex
\begin{table*}[t]
\caption{Quantitative ablation study of keywords and soft embeddings of preference conditions on two datasets. The best results are in \textbf{bold} and the second-best results are \underline{underlined}.}

\begin{tabular}{lcccccccc}
\toprule
Dataset             & \multicolumn{4}{c}{POG}                                                        & \multicolumn{4}{c}{MovieLens}                                                  \\ \cmidrule(lr){2-5}\cmidrule(lr){6-9} 
\multirow{2}{*}{Metric} & \multicolumn{2}{c}{LPIPS($\downarrow$)} & \multicolumn{2}{c}{SSIM($\uparrow$)} & \multicolumn{2}{c}{LPIPS($\downarrow$)} & \multicolumn{2}{c}{SSIM($\uparrow$)} \\ \cmidrule(lr){2-3}\cmidrule(lr){4-5} \cmidrule(lr){6-7}\cmidrule(lr){8-9} 
                    & History            & Target             & History           & Target            & History            & Target             & History           & Target           \\ \hline
\methodnameshort    & \textbf{0.5375}    & \textbf{0.5482}    & {\ul 0.1640}      & {\ul 0.1600}      & \textbf{0.4190}    & \textbf{0.4140}    & {\ul 0.2486}      & \textbf{0.2515} \\
\xspace\xspace w/o embeddings      & {\ul 0.5455}       & 0.5592             & \textbf{0.1652}   & \textbf{0.1608}   & {\ul 0.4215}       & {\ul 0.4176}       & \textbf{0.2488}   & {\ul 0.2505}     \\
\xspace\xspace w/o keywords        & 0.5616             & 0.5535             & 0.1533            & 0.1590            & 0.4406             & 0.4390             & 0.1867            & 0.1858           \\
\xspace\xspace w/o both & 0.5626             & {\ul 0.5526}       & 0.1531            & 0.1567            & 0.4561             & 0.4542             & 0.1589            & 0.1575           \\
\bottomrule
\end{tabular}

\label{tab:ablation_hardsoft}
\end{table*}

%% file: img/soft_prompt.tex
\begin{figure}[t]
  \centering
  \begin{minipage}[b]{0.23\textwidth}
    \centering
    \includegraphics[width=0.6\textwidth]{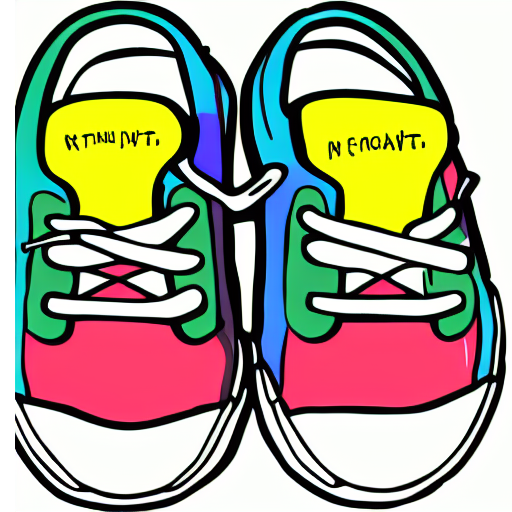}
    \subcaption{W/o soft embeddings.}
  \end{minipage}
  \hfill
  \begin{minipage}[b]{0.23\textwidth}
    \centering
    \includegraphics[width=0.6\textwidth]{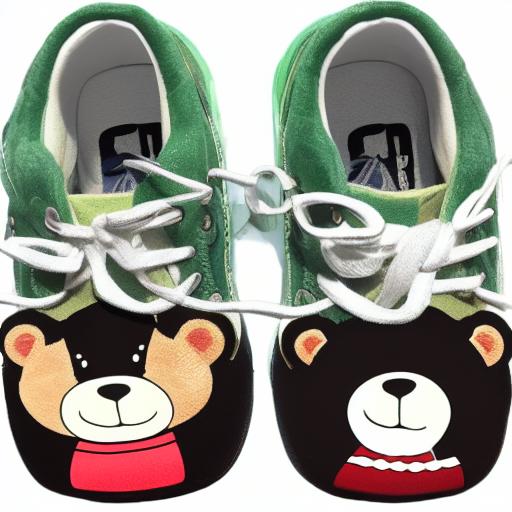}
    \subcaption{W/ soft embeddings.}
  \end{minipage}
  \caption{A case study of soft preference embeddings reveals the presence of language bias. With keywords ``shoes'' and ``cartoon'', the generation without these embeddings may produce a cartoon-style drawing of shoes, whereas the desired outcome is realistic shoes adorned with cartoon patterns.}
  \label{fig:embedding}
\end{figure}

%% file: table/ablation_finetune.tex
\begin{table}
\caption{Quantitative ablation study of P-tuning V2 and multimodal tokens using the LPIPS metric two datasets. $L$ denotes the number of multimodal tokens. The best results are in \textbf{bold} and the second-best results are \underline{underlined}.}
\begin{tabular}{ccccc}
\toprule
ID                  & P-Tuning V2                  & $L$                  & POG               & MovieLens        \\ \hline
1                   & \ding{55}                    & 2                    & 0.4398            & 0.5471           \\
2                   & \ding{55}                    & 4                    & 0.4353            & 0.5522           \\
3                   & \ding{55}                    & 8                    & 0.4421            & 0.5586           \\
4                   & \ding{55}                    & 16                    & 0.4482            & 0.5690           \\ \hline
5                   & \ding{51}                    & 2                    & 0.4230            & 0.5453           \\
6                   & \ding{51}                    & 4                    & {\ul 0.4190}      & \textbf{0.5375}  \\
7                   & \ding{51}                    & 8                    & \textbf{0.4155}   & {\ul 0.5386}     \\
8                   & \ding{51}                    & 16                    & 0.4212            & 0.5406           \\
\bottomrule
\end{tabular}
\label{tab:ablation_finetune}
\end{table}

%% file: table/auxiliary.tex
\begin{table}
\caption{Comparison of the recommendation performances between MMGCN leveraging different image features of items and users. The best results are in \textbf{bold} and the second-best results are \underline{underlined}.}
\begin{tabular}{ccccc}
\toprule
                        & Item             & User            & Recall@10                & NDCG@10               \\ \hline
No-image                & \ding{55}        & \ding{55}       & 17.57\%                  & 0.0859                \\
Item-only               & \ding{51}        & \ding{55}       & 18.88\%                  & 0.0947                \\
Averaged-user           & \ding{51}        & Average         & {\ul 19.54\%}            & {\ul 0.0989}          \\
Generated-user          & \ding{51}        & Generated       & \textbf{20.03\%}         & \textbf{0.1004}       \\
\bottomrule
\end{tabular}
\label{tab:auxiliary}
\end{table}

%% file: section/conclusion.tex
\section{Conclusion and Further Work}
In this paper, we have proposed a method named PMG for personalized multimodal generation using LLMs. By leveraging large language models, we extracted user preferences and used them to condition the generation process of a generator. The experiments on image generation validate the effectiveness of PMG and its potential for downstream recommendation tasks. This work paves the way for further advancements in personalized generation, enabling the creation of tailored and engaging user experiences.

In future work, we aim to enhance the realism of the generated images. We plan to employ retrieval-based augmentation by incorporating real image inputs as references to guide the generation of more realistic images, addressing the issue of hallucination.


%% file: section/appendix.tex
\begin{figure*}[t]
    \centering
    \includegraphics[width=0.95\linewidth]{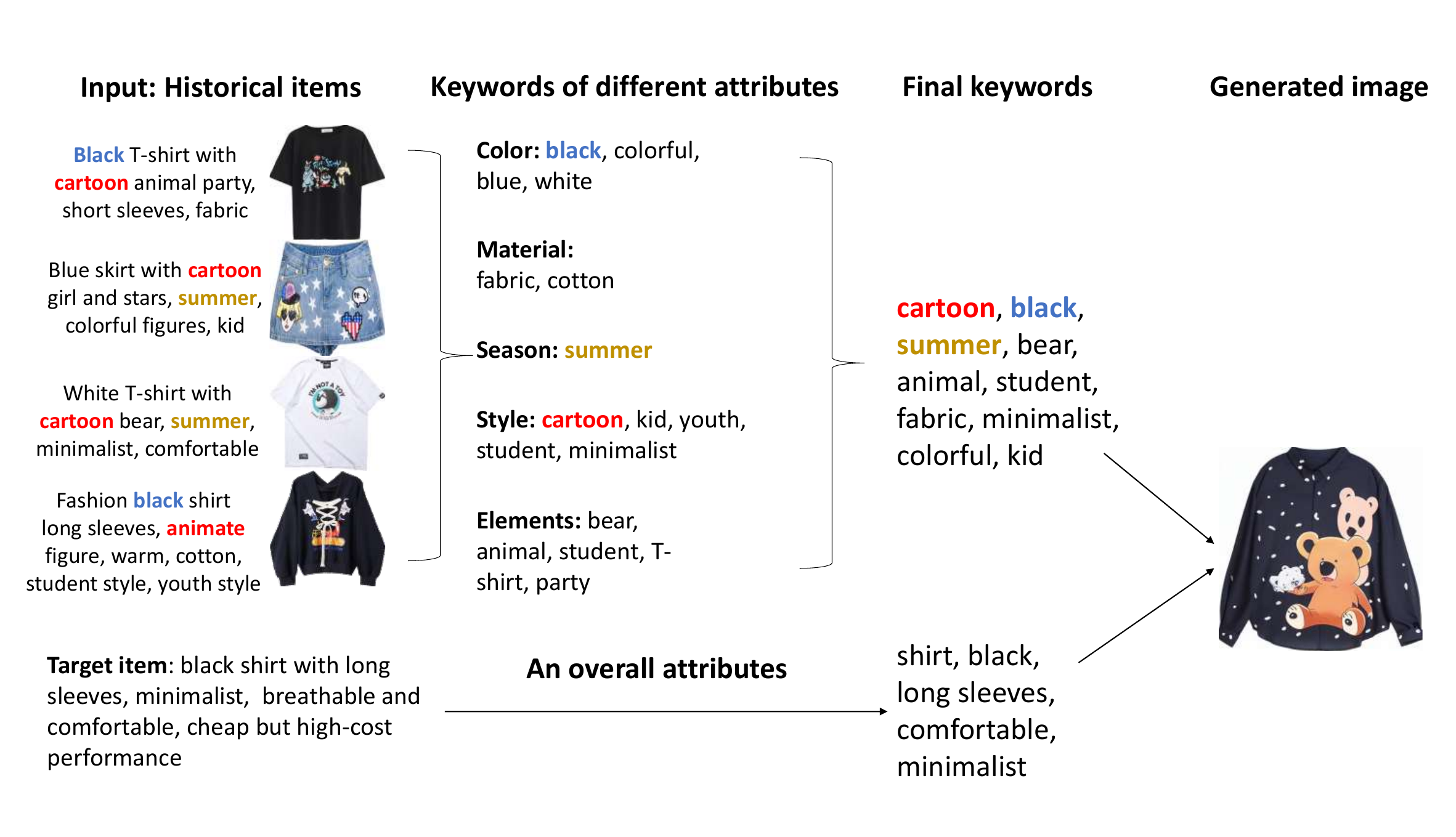}
    \caption{An example of keywords while generating a cartoon-style shirt.}
    \label{fig:keyword_example}
\end{figure*}

\section{Implementation Details}
\label{sec:Implement}
In all of our experiments\footnote{ \url{https://github.com/mindspore-lab/models/tree/master/research/huawei-noah/PMG}}, we select Llama2-7B~\cite{Llama2} as the basic LLM model and Stable Diffusion V1.5~\cite{stable_diffusion} as the image generator. Due to limitations of the dataset, at most $ n=10 $ historical items and only the current $ m=1 $ conversation are considered in the prompt of user preferences extraction. Then 10 personalized keywords and 5 target keywords are extracted for image generation because of the limitation of input. In the training of soft preference embeddings, $ L=4 $ multimodal tokens and $ S=4 $ prefix embeddings are used to get personal embedding. In our experiments, this training process costs 12 hours on a single NVIDIA V100 GPU with a learning rate of $10^{-5}$. As for inference, each image costs about 5 seconds, 2s for LLM and 3s for stable diffusion.

\section{Example of Prompts}

\lstset{
  breaklines=true,
  basicstyle=\footnotesize\ttfamily\SetTracking{encoding=*}{-40}\lsstyle,
  columns=fullflexible,
  showstringspaces=false,
  commentstyle=\color{gray},
  xleftmargin=2em,
  frame=single,
}
Taking the prompts used for movie posters as an example, we first generate descriptions of each movie using the two prompts below.

The prompt for \textbf{summarizing movies}:
\begin{lstlisting}
### Human: Here is a movie. Movie title "<title>". Movie introduction "<introduction>". Movie Genre: "<genres>". Please summarize this movie using one sentence within 30 words.
### Assistant: This movie
\end{lstlisting}

The prompt for \textbf{captioning movie posters}:
\begin{lstlisting}
### Human: Here is a movie poster <movie_poster>. Please caption it.
### Assistant: The caption of this poster is:
\end{lstlisting}

Second, we \textbf{generate user preference keywords} using the prompts below, and the <watching history> is replaced with the generated movie descriptions from the above step.

\begin{lstlisting}
### Principle: The assistant is helping the human to generate keywords of a movie lover's interests.
### Human: A movie lover watched some movies. Please provide 10 keywords to describe his movie interests especially on <attribute>. The example of output is "The keywords are: 1. Keyword 1; 2. Keyword 2; ..." His historical conversations are: <conversation history>. The movies he watched are: <watching history>.
### Assistant: The keywords are:
\end{lstlisting}

Third, we \textbf{generate the target item keywords} for the movie using the prompts below.

\begin{lstlisting}
### Human: Here is a movie. Movie title "<title>". Movie introduction "<introduction>". Movie Genre: "<genres>". Please describe this movie with 5 keywords. Keywords can be related to its genre, country, style or era.
### Assistant: The 5 keywords are:
\end{lstlisting}

Finally, the user preference keywords and target item keywords are used as input of the generator (prompt of stable diffusion in this paper) to generate multi-modal results.

The prompts of different scenarios are similar, just slightly adjusted to the application. The prompts of movies and clothes scenarios only differ in the words "watch", "movies" and "buy", "clothes". The prompts of emoticons are slightly different. Its target keywords are not descriptions of a specific item but moods reflected in the conversation and corresponding expressions or actions.

\section{Example of Keywords}

Figure~\ref{fig:keyword_example} is an example of keywords while generating a cartoon-style shirt (the same example as the one in Figure 4 of our paper). Conversation inputs and soft embeddings are omitted for brevity. Keywords of different attributes are generated first and then 10 of them are selected as the final preference keywords. 

\textbf{The description of the user behavior}, i.e., the historically clicked items are:
\begin{lstlisting}
1. Black T-shirt with cartoon animal party, short sleeves, fabric
2. Blue skirt with cartoon girl and stars, summer, colorful figures, kid
3. White T-shirt with minimalist cartoon bear, summer, comfortable
4. Fashion black shirt, long sleeves, animate figure, warm, cotton, student style, youth style
\end{lstlisting}

\textbf{The keywords for the attributes} “Color”, “Material”, “Season”, “Style” and “Elements” are learned from the user behaviors and listed below. In the above 4 clicked items by the user, “cartoon” appears in all of them and hence is an important keyword, and we can see that “cartoon” is successfully learned as a user preference in the attribute of “Style” below.
\begin{lstlisting}
Color: black, colorful, blue, white
Material: fabric, cotton
Season: summer
Style: cartoon, kid, youth, student, minimalist
Elements: bear, animal, student, T-shirt, party
\end{lstlisting}

Among all the above keywords for user preference, we choose the top-10 and obtain \textbf{the final user preference keywords} as follows:

\begin{lstlisting}
1. cartoon  2. black    3. summer       4. bear     5. animal
6. student  7. fabric   8. minimalist   9. colorful 10. kid
\end{lstlisting}

For a target item given by the recommender system, we already have the item’s description (e.g., the black shirt below). From the description, we use the LLM to \textbf{summarize the keywords} as below.

\begin{lstlisting}
Description: black shirt with long sleeves, minimalist, breathable and comfortable, cheap but high-cost performance
Generated target item keywords: shirt, black, with long sleeves, comfortable, minimalist
\end{lstlisting}

Finally, the user preference keywords, target item keywords, and soft preference embeddings obtained above are used for the image generation.